\documentclass[twocolumn]{aastex631}

\usepackage{amsmath}	
\usepackage{multirow}

\begin{document}

\title{SgrA$^{*}$ spin and mass estimates through the detection of multiple extremely large mass-ratio inspirals}

\author[0000-0002-9458-8815]{Ver\'{o}nica V\'{a}zquez-Aceves}
\affiliation{Kavli Institute for Astronomy and Astrophysics at Peking University, 100871 Beijing, China}

\author[0000-0003-2509-6558]{Yiren Lin}
\affiliation{Astronomy Department, School of Physics, Peking University, Beijing 100871, China}

\author[0000-0002-5467-3505]{Alejandro Torres-Orjuela}
\email{Corresponding author: atorreso@bimsa.cn}
\affiliation{Beijing Institute of Mathematical Sciences and Applications, Beijing 101408, China}


\begin{abstract}

We analyze the parameter estimation accuracy that can be achieved for the mass and spin of SgrA$^\ast$, the SMBH in our Galactic Center, by detecting multiple extremely large mass-ratio inspirals (XMRIs). XMRIs are formed by brown dwarfs (BD) inspiraling into a supermassive black hole (SMBH), thus emitting gravitational waves (GWs) inside the detection band of future space-based detectors such as LISA and TianQin. Theoretical estimates suggest the presence of approximately 10 XMRIs emitting detectable GWs, making them some of the most promising candidates for space-based GW detectors. Our analysis indicates that even if individual sources have low SNRs ($\approx10$), high-precision parameter estimates can still be achieved by detecting multiple sources. In this case, the accuracy of the parameter estimates increases by approximately one to two orders of magnitude, at least. Moreover, by analyzing a small sample of 400 initial conditions for XMRIs formed in the Galactic Center, we estimate that almost 80\,\% of the detectable XMRIs orbiting SgrA$^\ast$ will have eccentricities between 0.43 to 0.95 and an $\mathrm{SNR}\in [10,100]$. The remaining $\sim$20\,\% of the sources have an $\mathrm{SNR}\in [100,1000]$ and eccentricities ranging from 0.25 to 0.92. Additionally, some XMRIs with high SNR are far from being circular. These loud sources with $\mathrm{SNR}\approx 1000$ can have eccentricities as high as $e\approx0.7$; although their detection chances are low, representing $\lesssim$2\,\% of the detectable sources, their presence is not ruled out.

\end{abstract}


\section{Introduction}

Extremely large mass ratio inspirals (XMRIs) are inspiraling systems with a mass ratio of $q\sim 10^8$ composed of a brown dwarf (BD) and a supermassive black hole (SMBH)~\citep{Pau_2019,gourgoulhon_le-tiec_2019} that emit gravitational waves (GWs) within the detection band of space-based GW detectors such as LISA~\citep{LISA_2017,lisa_2024} or TianQin~\citep{TianQin_2016, TianQin_2021,tianqin_2024}. Their long merger timescales of about $10^8\,{\rm yr}$ allow these sources to accumulate within their host systems; this distinctive characteristic makes their detection highly promising. Moreover, their formation and evolution models indicate that XMRIs emitting GWs are currently present at the center of our Galaxy~\citep{Pau_2019}, placing them among the closest sources that the space-based detectors will detect. In our previous work~\citep{VVA_2023}, we derived the number of circular and eccentric XMRIs emitting detectable GWs for different values of the spin of SgrA$^\ast$ and obtained the accuracy that can be achieved in the measurements of the spin and mass of SgrA$^\ast$ if a single source with high signal-to-noise ratio (SNR) of about 30 to 2000 is detected.

In this work, we expand our approach to include a more conservative scenario by considering the simultaneous detection of multiple lower-SNR XMRIs ($\mathrm{SNR}\lesssim200$) to perform the accuracy analysis. We generate a set of 400 orbital parameters for XMRIs and randomly select systems with an SNR within two specified ranges, regardless of the orbital parameters. The first range corresponds to sources with a low SNR of around 20, while the second SNR range represents moderate SNR values, $\mathrm{SNR}\approx100$. We generate the waveforms of the XMRIs using a model developed for high mass ratio sources, including Post Newtonian (PN) corrections up to an order of 2.5~\citep{Barack:2003fp}. Furthermore, we perform a Fisher matrix analysis~\citep{coe_2009} assuming five sources are detected simultaneously but with distinguishable signals to estimate the accuracy of measuring the spin and mass of SgrA$^\ast$. The Fisher information matrix captures how sensitive the GW signal is to various parameters, quantifying the expected errors in their measurement; it assumes Gaussian noise, making it suitable for a first-order approximation of uncertainties in parameter estimation.

In Section~\ref{sec:2}, we describe the properties of the inspiraling systems we focus on: XMRIs formed in the Galactic Center orbiting SgrA$^\ast$. In Section~\ref{sec:Phasespace}, we obtain the number of XMRIs expected to orbit around SgrA$^\ast$ with different SNR, showing that almost 80\,\% of the detectable XMRIs will have an $\mathrm{SNR}\in [10,100]$ and about 20\,\%, an $\mathrm{SNR}\in [100,1000]$. We give their eccentricity range and show that only about 2\,\% of the detectable sources will have higher SNRs above 1000. In Section~\ref{sec:Accuracy}, we show that by detecting multiple XMRIs, the accuracy of parameter estimation can be enhanced. The accuracy for the spin, $\Delta s$, and the mass, $\Delta M$, increases with N, the number of detected XMRIs. We present our results for the Fisher matrix analysis considering up to N=5 simultaneous XMRIs and discuss our results to finally give our conclusions in Section~\ref{sec:Conclusions}.

\section{Galactic XMRIs}
\label{sec:2}

Models of galactic density distributions~\citep{BW_1977} and two-body relaxation processes suggest that XMRIs can form~\citep{Sigurdsson_1997, Hopman_2005, Amaro_2007, Pau_2019, Pau_2020} in the center of galaxies, and detecting them within our Galaxy provides a unique opportunity to test and refine these models. The results obtained can then be applied to other galaxies with similar characteristics to get a general understanding of their structure and dynamics. However, if it is not via GWs, the direct detection of XMRIs might not be achievable in our galactic center, and even less so in other galaxies. Therefore, these systems are among several difficult-to-observe phenomena we expect to reveal with future space-based GW detectors. Detecting the GW emission of galactic XMRIs will, furthermore, enable us to map the spacetime around SgrA$^\ast$, offering unprecedented insight into the properties of the SMBH. In particular, it will help us determine the magnitude of SgrA$^\ast$'s spin, a parameter that is difficult to measure using electromagnetic observations.

We focus on XMRIs formed via two-body relaxation processes at the center of our galaxy. According to our previous estimates~\citep{VVA_2023} and those presented by~\cite{Pau_2019}, there is a population of $N\approx15$ XMRIs around SgrA$^\ast$ emitting GWs in the band of space-based detectors right now, with eccentric sources $(e> 0.8)$ being more abundant ($N\approx$8-15) than circular sources with eccentricities $e\lesssim 0.2$ ($N\approx$1-5). These XMRIs have random orbital parameters, yet all have a semimajor axis that is smaller than the critical semimajor axis, $a_{\rm crit}$, required to form an inspiraling system~\citep{Hopman_2005, Pau_2020}.

The value of $a_{\rm crit}$ is computed by comparing the merger timescale $T_{\rm GW}$ with the time it takes the relaxation processes to change the pericenter of the orbit; this timescale is given by the angular momentum diffusion timescale $T_{\rm rlx}\times(1-e^2)$ \citep{Hopman_2005}. It is assumed that within the influence radius of SgrA$^\ast$, approximately 3\,pc, the dominant driver of relaxation is the population of stellar-mass black holes (BHs) with typical masses of $m_{\rm BH} \simeq 10\,{\rm M_\odot}$. These BHs sink to the center via mass segregation, forming a density cusp; less massive objects -- main-sequence (MS) stars, neutron stars (NSs), white dwarfs (WDs), and brown dwarfs -- settle into shallower profiles and do not significantly impact the relaxation processes.

\cite{BW_1977} predicted that objects around a SMBH distribute into a cusp of the form $\rho(r)\propto r^{-\gamma}$, with heavier objects sinking deeper into the potential well due to energy equipartition, leading to steeper cusps for more massive species: $\gamma=1.75$ for BHs and $\gamma_0=1.5$ for less massive objects such as MS. Subsequent numerical simulations and observations confirmed this scenario; for example, \cite{Marc_2006} finds $\gamma_0$ falls between 1.3 and 1.4, while in the work of \cite{Hopman_2006, AlexanderTal_2009} strong mass segregation occurs when the number of heavy objects is relatively low, creating a steeper cusp with $\gamma\in (2,2.75)$ and $\gamma_0 \in (1.5, 1.75)$.

The density distribution of objects can affect the orbital parameters of the inspiraling systems as well as their event rates. For XMRIs adopting a Bahcall-Wolf (BW) cusp leads to event rates of $\sim 10^{-6}$~yr$^{-1}$, while in the strong mass segregation scenario, it increases by an order of magnitude, $\sim 10^{-5}$~yr$^{-1}$ \citep{Pau_2019}. However, detailed observations and N-body simulations performed by \cite{Gallego-Cano_2018, Schodel_2018, Baum_2018} show that in our Galaxy stars distribute in a cusp with $\gamma_0 \in (1.13 -1.43)$, which points to a non-strong mass segregation scenario. Also, the detailed measurements of S2's orbit reported by \cite{GravityColl_2024} show good agreement with a dynamically relaxed stellar cusp in the Galactic Center.

Based on these results, we assume that BDs follow the MS distribution and take the BW values, $\gamma=1.75$ and $\gamma_0=1.5$, as a good approximation for the density distribution of the objects in the center of the Galaxy. Our estimates represent a lower limit, as for galaxies where strong mass segregation occurs, the event rates of inspiraling systems can be enhanced~\citep{Preto_2010, Amaro-Seoane_2011}.

The semimajor axis $a$, and pericenter $r_{\rm p} = a (1-e)$, of newly formed XMRIs must satisfy $a\lesssim a_{\rm crit}$ and $r_{\rm p} > r_{\rm tidal}$, where $r_{\rm tidal}$ is the tidal radius; otherwise, the BD will be disrupted by the tidal forces of the SgrA$^\ast$. Additionally, $r_{\rm p}$ also must lie outside the position of the last stable orbit (LSO), $r_{\rm LSO}$, or the BD will plunge into the SMBH without performing an inspiraling process when reaching the pericenter. This defines a maximum eccentricity
$e \lesssim e_{\rm max}$ given by
\begin{equation}\label{eq:emax}
    e_{\rm max}=1-\frac{{\rm R_L}}{a};
\end{equation}
where ${\rm R_L}$ is the maximum value between $r_{\rm LSO}$ and $r_{\rm tidal}$. Assuming a BD with mass $m_{\rm BD}=0.05\,{\rm M_\odot}$ and radius $r_{\rm BD}=0.083\,{\rm R_\odot}$~\citep{Sorahana_2013}, we get
\begin{equation}\label{eq:rtidal}
    r_{\rm tidal}= \left( 2 \frac{(5-n) M_{\rm  SgrA^\ast}}{3 m_{\rm BD}}\right)^{1/3} r_{\rm BD} \approx 2.8\,{\rm R_S},
\end{equation}
where we used a polytropic index $n=1.5$~\citep{shapiro_teukolsky_1983}. 

The position of the LSO depends on the spin magnitude of the SMBH and the orbital inclination, $\iota$, of the inspiraling object's orbit with respect to the spin axis. An inspiraling system in a prograde orbit around a highly spinning SMBH can reach smaller pericentre distances and higher eccentricities than one formed around a slowly spinning SMBH. The influence that the spin of the SMBH has on the formation of GW sources as well as on their event rates was studied by~\cite{Pau_2013}. This effect applies to all inspiraling systems, so the spin magnitude also influences the event rates and the initial orbital parameters of an XMRI. We obtain the position of the LSO of a Kerr black hole as ${\rm R_{LSO}} = 4 {\rm R_S} \mathcal{W}(\iota,s)$, where ${\rm R_S}=2GM_{\rm  SgrA^\ast}/c^2$ is the Schwarzschild radius, $G$ is the gravitational constant, $c$ is the speed of light in vacuum, and $\mathcal{W}(\iota, s)$ is a function, introduced in~\cite{Pau_2013}, that accounts for effect of the spin magnitude and the orbital inclination of the inspiraling object.

To get the value of $a_{\rm crit}$ we solve $T_{\rm GW}<T_{\rm rlx} \times (1-e^2)$ for $a_{\rm crit}$. With the merger timescale \citep{Peters:1964} encoding the orbital properties of the inspiraling systems written as
\begin{equation}
    T_{\rm GW} = \sqrt{2}~ \frac{25}{85}\frac{a_0^4~c^5}{G^2 m_{\rm BD} M_{\rm  SgrA^\ast}^2} ~(1-e_0)^{7/2}.
\end{equation}
The influence of the density distribution model and the assumption that BHs lead the relaxation processes inside the influence radius, which we take as $R_{\rm h}= 3pc$, is embedded in the $T_{\rm rlx}$ term. With these factors in consideration, the critical semimajor axis can be written as
\begin{align}
    a_{\rm crit} &= R_{\rm h} \left[ \mathcal{W}(\iota,s) N_0 \ln(\Lambda) \frac{m_{\rm BH}^2}{M_{\rm  SgrA^\ast} ~m_{\rm BD}} \times \epsilon \right]^{\frac{1}{\gamma-3}}, 
    \label{eq:a_crit}
\end{align}
where $\epsilon \simeq 17~ (3-\gamma)(1+\gamma)^{3/2}$, $N_0=10^4$ is the number of BHs within $R_{\rm h}$, $\ln(\Lambda) = 13$ is the Coulomb logarithm \citep{Binney_Tremaine_1987}, and $\gamma = 1.75$.

For BDs, the average critical semimajor axis is $a_{\rm crit} \approx 10^{-3}\,{\rm pc}$. However, the parameters of inspiraling systems are also influenced by the spin magnitude of the SMBH and the orbital inclination of each orbit~\citep{Pau_2013}. We set SgrA$^\ast$ to be located at $8\,{\rm kpc}$, to have a mass of $M_{\rm  SgrA^\ast}=4\times10^6\,{\rm M_\odot}$~\citep{Ghez_2008, Gillessen_2009}, and its spin to be aligned with the line-of-sight following the results of the Event Horizon Telescope~\citep{SgrA1_2022, SgrA2_2022, SgrA3_2022, SgrA4_2022, SgrA5_2022, SgrA6_2022}. These results seem to indicate that the spin of SgrA$^\ast$ is pointing at a small angle ($< \pi/6\,{\rm rad}$) towards Earth and that the spin magnitude is $s > 0.5$. However, estimates for the spin magnitude of SgrA$^\ast$ remain controversial. Based on dark spots identified in Event Horizon Telescope images, \cite{Dokuchaev_2023} estimates the spin magnitude to range between 0.65 and 0.9, while~\cite{Daly_2024} gives an estimate of 0.87 to 0.90 based on X-ray and radio data. These results differ significantly from those of~\cite{Fragione_2020, Fragione_2022}, who estimate $s\lesssim0.1$ based on the spatial distribution of the S-stars. Therefore, our analysis considers spin magnitudes of $s=0.1$ and $s=0.9$; this also gives a lower and upper limit for the accuracy estimates if the spin magnitude takes an intermediate value.

\section{Phase space and signal-to-noise ratio}
\label{sec:Phasespace}

The extremely large mass-ratio of XMRIs, $q\sim 10^8$, provides a few advantages in comparison with other inspiraling systems: the motion of the BD can be approximated as being nearly geodesic, and the merger timescale is usually long $T_{\rm GW}\approx 10^8$ years. This long evolution period implies that the systems (1) can accumulate, giving rise to the estimates for the number of XMRIs emitting GWs in our Galactic Center at this moment, and (2) the frequency of the emitted GWs remains nearly constant during the detection period of the space-based detectors. These characteristics allow for a more straightforward computation of the waveforms, which, for the accuracy analysis, we generate by implementing PN corrections up to 2.5 following \cite{Barack:2003fp} and \cite{Fang_Huang_2020}. Moreover, we estimate the XMRIs SNR as in \cite{Barack:2003fp,finn_thorne_2000}
\begin{equation}\label{eq:snr_estimate}
    (\mathrm{SNR})^2=\int\frac{h_{c,n}^2}{5fS_n(f)}~\mathrm{d}\ln f
\end{equation}
where $h_{c,n}$ is the characteristic strain, $S_n$ is the noise spectral density, and the factor of 5 comes from sky-averaging. 

The initial semimajor axis and eccentricity of an inspiraling system are always assumed to be such that the pericenter lies outside R$_{\rm L}=$max($r_{\rm LSO}, r_{\rm tidal}$) to ensure that the object will not plunge or be tidally disrupted at the pericenter. For this reason, the typical values for $a$ and $e$ are obtained by assuming $a\approx a_{\rm crit}$, (cf. Equation~\ref{eq:a_crit}), and $e\approx e_{\rm max}$. However, for each $a\lesssim a_{\rm crit}$, there is a range of eccentricities that an XMRI can take as long as $T_{\rm GW}< T_{\rm rlx}\times(1-e^2)$~\citep{Pau_2018}. Following the description provided in our previous work~\citep{VVA_2023}, we sample the phase space and obtain a set of 200 possible orbital parameters for different orbital inclinations, $\iota=[0.1,0.4,0.7,1.0,1.57]\,{\rm rad}$, per spin magnitude; thus, in total, we obtained a sample of 400 initial orbital parameters which will be used to obtain an average of the number of detectable sources with $\mathrm{SNR}\approx$10, 100 and 1000.

This procedure gives a more direct description of the number of detectable sources we expect to detect as soon as the space-based detectors start operating. We approximate the SNR (Equation~(\ref{eq:snr_estimate})) as
\begin{equation}\label{eq:snr_approx}
    (\mathrm{SNR})^2
    \approx \sum_n\frac{h_{c,n}^2\dot{f_n}}{5f_n^2S_n(f)}T_\mathrm{obs},
\end{equation}
where $T_{\rm obs}$ is the observation time. However, due to the highly eccentric nature of XMRIs, it is computationally expensive to compute $h_{c,n}$; so for this part of the study in which we obtain the average number of sources with a given SNR, we compute $h_{c,n}$ as in Equation~(56) from \cite{Barack:2003fp}
\begin{equation}
    h_{c,n} = \frac{1}{\pi D} \sqrt{2\frac{\dot{E}_n}{\dot{f}_n}},
    \label{eq:h_En}
\end{equation}
where $D=8\,{\rm kpc}$ is the luminosity distance and $\dot{E}_n$ is the power radiated by GWs at the frequency $f_n$, given in Equations~(57) and~(58) in \cite{Barack:2003fp}. Although it still requires the computation of several harmonics, it allows for a quick estimate of the SNR. We take $n=500$ and use the following two equations
\begin{align}
    \frac{de}{dt} =& -\frac{e}{15} \frac{m_{\rm BD}}{M_{\rm  SgrA^\ast}^2}\frac{(2\pi M_{\rm  SgrA^\ast} f_{\rm orb})^{8/3}}{(1-e^2)^{5/2}} \left[ 304 + 121e^2 \right], \label{eq:dedt} \\
    \frac{df_{\rm orb}}{dt} =& \frac{96}{10\pi} \frac{m_{\rm BD}}{M_{\rm  SgrA^\ast}} \frac{(2\pi M_{\rm  SgrA^\ast})^{11/3}}{(1-e)^{7/2}} \left[ 1 + \frac{73}{24} e^2 + \frac{37}{96}e^4 \right], \label{eq:dfdt}
\end{align}
where $f_{\rm orb}$ is the orbital frequency, to evolve the 400 sources up to when Equation~(\ref{eq:snr_approx}) gives $\mathrm{SNR}\approx10$, $\mathrm{SNR}\approx100$, and $\mathrm{SNR}\approx1000$ assuming an observation time of two years.

This estimate is only valid because the frequency of XMRIs, i.e., its orbital parameters, do not change significantly during $T_{\rm obs}$, and $\dot{f}_n \approx n\dot{f}_{\rm orb}$. So Equation~(\ref{eq:snr_approx}) can be written in terms of the orbital frequency $f_{\rm orb}$ and for a given value of SNR, together with Equations~(\ref{eq:dedt}) and~(\ref{eq:dfdt}), we can obtain the corresponding orbital parameters ($a_{10_i}, e_{10_i}$), ($a_{100_i}, e_{100_i}$), and ($a_{1000_i}, e_{1000_i}$), where the subindex indicates the SNR of the source and $i\in[1,400]$.

Note that we only use Equation~(\ref{eq:h_En}) to obtain $h_{c,n}$ for the estimation of the orbital parameters that will result in an SNR with a specific value. For the estimates of the accuracy in Section~\ref{sec:Accuracy}, we obtain the waveform by evolving the orbital parameters with a PN formalism up to 2.5 PN based on \cite{Barack:2003fp} and obtain the SNR from the waveform with Equation~(\ref{eq:snr_estimate}).

With the orbital parameters, ($a_{10_i}, e_{10_i}$), ($a_{100_i}, e_{100_i}$), and ($a_{1000_i}, e_{1000_i}$), we apply the description given by \cite{Pau_2019} to estimate the number of sources distributed between given values of semimajor axes to obtain an average number of XMRIs orbiting SgrA$^\ast$ for each value of the SNR. The total number of sources orbiting around ${\rm SgrA^\ast}$ is
\begin{equation}\label{eq:N_tot}
    N_{\rm XMRIs}= \dot{\Gamma}_{\rm XMRI} (a_{\rm crit})  \times  T_{\rm GW},
\end{equation}
where $\dot{\Gamma}_{\rm XMRI}$ is the XMRI event rate, given by
\begin{equation}\label{eq:EventRate}
    \dot{\Gamma}_{\rm XMRI} = \frac{3}{5 T_0}\frac{ N_{\rm IR}}{\text{R}_{\rm h}^{\lambda}} f_{\rm sub}\left\lbrace\left[ a_{\rm crit}^{\lambda}\left(\text{ln}\left(\frac{a_{\rm crit}}{2 \text{R}_{\rm L}}\right)- \frac{1}{\lambda}\right)\right] \right\rbrace,
\end{equation}
with $\lambda=9/2 - \gamma_0 - \gamma$, $N_{\rm IR} = M_{\rm  SgrA^\ast} / \bar{m}$ the number of objects within the influence radius of SgrA$^\ast$, ${\rm R_h}= 3\,{\rm pc}$, $T_0\approx 6.9\times10^9\,{\rm yr}$ a normalization timescale, $\bar{m} = 0.27\,{\rm M_\odot}$ the average stellar mass, $f_{\rm sub}=0.21$ the fraction of BDs obtained from a Kroupa broken power law~\citep{Kroupa_2001}, and ${\rm R_L}= \max(r_{\rm LSO},r_{\rm tidal})$.

The estimates given by Equation~(\ref{eq:N_tot}) need to be refined, as this total number accounts for all XMRIs formed, regardless of their GW frequency. Although a newly formed XMRI is emitting GWs, the emission is not within the detection band of future space-based detectors; an XMRI evolves due to the energy loss by GWs, shrinking its semimajor axis and circularizing. Therefore, by implementing a continuity function to describe the evolution of the sources in the phase space, we can estimate the number of XMRIs at a given point in their evolution
\begin{equation}
    \frac{\partial}{\partial a} (\dot{a}(a,e)~g) ~+ ~\frac{\partial g}{\partial t} =0,
    \label{eq:continu}
\end{equation}
where $g=dN/da$ is the density function of the number of sources $N$ per semimajor axis $a$, and 
$\dot{a}(a,e)$ is the evolution of the semimajor axis due to the emission of GWs \citep{Peters:1964}.

In the original description~\citep{Pau_2019}, the author obtains the number of eccentric and circular XMRIs by setting the following limits for the orbital parameters: $\kappa =\left[(a_{\rm crit}, e_{\rm crit}), (a_{\rm band}, e_{\rm band}), (a_{\rm break}, e_{\rm break}),(a_{\rm min}, e_{\rm min})\right]$, where $(a_{\rm crit}, e_{\rm crit})$ are the initial orbital parameters for a newly formed XMRI and $(a_{\rm min}, e_{\rm min})$ the merger parameters, defining the phase space over which an XMRI evolves. The number of detectable sources is computed at two intermediate stages: when the system enters the band of space-based detectors at $(a_{\rm band}, e_{\rm band})$, typically still highly eccentric ($e \gtrsim 0.9$), and when it circularizes at $(a_{\rm break}, e_{\rm break})$, with $e_{\rm break} \simeq 0.2$. We apply the same description to compute the number of detectable sources with a given SNR independently of their orbital parameters with the assumptions described in the previous sections, namely, the stellar mass black holes, $m_{\rm BH} = 10\,{\rm M_\odot}$, dominate the relaxation processes within the influence radius of SgrA$^\ast$, $R_{\rm h} = 3$ pc, and the objects follow a BW density profile $\gamma= 1.75$ for the BHs, and $\gamma_0 =1.5$ for the BDs.

We take the 400 representative sets of orbital parameters, $\kappa_i=\left[(a_{\rm crit_i}, e_{\rm crit_i}), (a_{10_i}, e_{10_i}), (a_{100_i}, e_{100_i}), (a_{1000_i}, e_{1000_i})\right]$ and obtain the number of sources $N_{10_i}, N_{100_i}$ and $N_{1000_i}$ for each set $\kappa_i$. We then proceed to average the number of sources obtained for each range of SNR. The total number of detectable sources is consistent with previous results~\citep{Pau_2019, VVA_2023}. We obtain an average of $N_{\rm tot} = 16^{+3}_{-5}$ for a spin magnitude of $s=0.9$ and of $N_{\rm tot} = 9^{+6}_{-4}$ for $s=0.1$; showing that despite variation in initial conditions, the evolution of XMRIs across a broad phase space converges to similar detection rates. The wide sampling allows us to capture the different orbital configurations that contribute to different SNR regimes: sources with $\mathrm{SNR}\approx10$, which represent 76\,\% of the total number of detectable sources, can take eccentricities ranging between 0.5 to 0.98, while sources with $\mathrm{SNR}\approx100$ that represent 22\,\% of the total number of sources, can have eccentricities between $0.21 \lesssim e\lesssim0.95$. Finally, loud sources with $\mathrm{SNR}\approx 1000$ and an eccentricity range $0\lesssim e\lesssim 0.73$, represent just 2\,\% of the total number of detectable sources. Table \ref{tab:Nresults} summarizes these results and shows that not all the detectable sources are highly eccentric, highlighting the fact that future searches for XMRI GW signals should not focus only on highly eccentric systems.

\begin{table}
    \centering
    \begin{tabular}{c|c|c|c}
        \hline
         SNR & s=0.1 &  s=0.9 & ${\rm N_{SNR}}/{\rm N_{tot}}$ \\
        \hline
           $\gtrsim10$   & $e\in$(0.51 - 0.97) &  $e\in$(0.43 - 0.98)  & 0.76  \\ \hline
           $\approx100$  & $e\in$(0.25 - 0.92)&   $e\in$(0.21 - 0.95)  & 0.22 \\ \hline
           $\gtrsim1000$ & $e\in$(0.11 - 0.67) &   $e\in$(0.09 - 0.73)  & 0.02  \\ \hline
    \end{tabular}
    \caption{The eccentricity range and ratio ${\rm N_{SNR}}/{\rm N_{tot}}$ for each spin magnitude, where ${\rm N_{SNR}}$ is the number of sources with an average SNR of $\gtrsim10$, $\approx100$, and $\gtrsim1000$ obtained from the set of 400 initial orbital parameters.}
    \label{tab:Nresults}
\end{table}

\section{Accuracy of mass and spin estimates for S\lowercase{gr}A$^\ast$ with multiple XMRIs}
\label{sec:Accuracy}

We evaluate the precision with which the spin and mass of SgrA$^\ast$ can be determined by conducting a Fisher matrix analysis~\citep{coe_2009}. This method provides a linearized approximation for the measurement uncertainties and approaches the actual errors in the high-SNR regime. The Fisher matrix for an individual XMRI can be expressed as
\begin{equation}\label{eq:deffm}
    (\Gamma_N)_{i,j} := \left\langle\frac{\partial h(\boldsymbol{\theta}_N)}{\partial\theta_i},\frac{\partial h(\boldsymbol{\theta}_N)}{\partial\theta_j}\right\rangle,
\end{equation}
where $\boldsymbol{\theta}_N := (\theta_{N,1},...,\theta_{N,n})$ is the vector corresponding to the $N$-th XMRI in the $n$-dimensional parameter space and $\langle\cdot,\cdot\rangle$ is the noise-weighted inner product~\citep{finn_1992,klein_barausse_2016}. As the harmonics of the different XMRIs almost do not evolve during the observation time, we assume their frequencies can be separated clearly during detection. Using this fact together with the linearity of the Fisher matrix, the Fisher matrix of the combined signal takes the form~\citep{isoyama_nakano_2018,torres-orjuela_huang_2023}
\begin{equation}\label{eq:totfm}
    \Gamma_\text{tot} = \sum_N \Gamma_N.
\end{equation}

To perform the accuracy analysis, we consider 20 random sources in total, separated into two sets of 10 sources per spin magnitude, $s=[0.1,0.9]$. From the set of 400 sources, we take five sources with $\mathrm{SNR}\approx20$, and five more with $\mathrm{SNR}\approx100$. The lower limit in this range represents sources above the detection threshold\footnote{We use the detection threshold of extreme mass ratio inspirals here as they are the most similar sources while there is no comparable estimate for XMRIs.}~\citep{babak_gair_2017}, while the upper limit represents a moderate SNR value well above the detection threshold. A deeper study would involve a full posterior sampling considering all the parameters of the source~\citep{li_2013,ye_fan_2023}, but we restrict ourselves to a Fisher matrix analysis because its lower computational costs allow us to explore the combination of multiple sources more easily.

We obtain the waveforms of the sources considering up to 2.5 PN corrections~\citep{Barack:2003fp}, but to reduce the computational cost of calculating them, we first use Equation~\ref{eq:snr_approx} to estimate the contribution of each harmonic and select the harmonics that contribute in total 99\,\% of the emitted power. The considered harmonics and orbital parameters for each of the 20 selected sources are shown in Table~\ref{tab:parameters}; the actual SNR of each source is computed from the waveform using Equation~\ref{eq:snr_estimate} and is shown in the right panel of the same table.

\begin{table}
    \centering
    \begin{tabular}{c|c|c|c|c|c}
        \hline
        $s$ & $\iota$ & $e$ & $r_\mathrm{p}[r_\mathrm{S}]$ & $n_\mathrm{h}$ & SNR \\
        \hline
        \multirow{5}{*}{0.1}
        & 0.1 & 0.852888 & 7.53890 & 19-118 (100)   & 29.7 \\
        & 0.4 & 0.717905 & 8.25989 & 8-42 (35)      & 28.0 \\
        & 0.7 & 0.523073 & 8.79724 & 1-20 (20)      & 25.0 \\
        & 1.0 & 0.907250 & 7.02318 & 38-237 (200)   & 17.5 \\
        & 1.57& 0.567448 & 8.71449 & 1-25 (25)      & 11.0 \\
        \hline
        \multirow{5}{*}{0.9}
        & 0.1 & 0.961463 & 6.09090 & 143-863 (719)  & 30.0 \\
        & 0.4 & 0.646565 & 8.50378 & 4-33 (30)      & 28.2 \\
        & 0.7 & 0.567355 & 8.70454 & 1-25 (25)      & 24.3 \\
        & 1.0 & 0.767857 & 8.05733 & 10-59 (50)     & 17.6 \\
        & 1.57& 0.439907 & 8.96971 & 1-18 (18)      & 11.6 \\
        \hline\\
        \hline
        \multirow{5}{*}{0.1}
        & 0.1 & 0.506323 & 6.28613 & 1-25 (25)      & 169 \\
        & 0.4 & 0.370228 & 6.48041 & 1-16 (16)      & 172 \\
        & 0.7 & 0.258564 & 6.62610 & 1-15 (15)      & 163 \\
        & 1.0 & 0.604174 & 6.10792 & 1-35 (35)      & 96 \\
        & 1.57& 0.279065 & 6.59283 & 1-15 (15)      & 73 \\
        \hline
        \multirow{5}{*}{0.9}
        & 0.1 & 0.769921 & 5.64530 & 9-55 (47)      & 152 \\
        & 0.4 & 0.581592 & 6.15145 & 1-30 (30)      & 148 \\
        & 0.7 & 0.280562 & 6.60106 & 1-12 (12)      & 149 \\
        & 1.0 & 0.410115 & 6.43219 & 1-20 (20)      & 105 \\
        & 1.57& 0.316318 & 6.55430 & 1-15 (15)      & 70 \\
        \hline
    \end{tabular}
    \caption{The XMRIs considered for the accuracy analysis: For each system, we show (from left to right) the spin magnitude of SgrA$^\ast$ $s$, the orbital inclination $\iota$, the eccentricity $e$, the pericenter distance $r_\mathrm{p}$, the range of harmonics calculated (with the total number of harmonics in the parenthesis), and the SNR obtained from the noise-weighted inner product.}
    \label{tab:parameters}
\end{table}

\begin{figure}
    \centering
    \includegraphics[width=\linewidth]{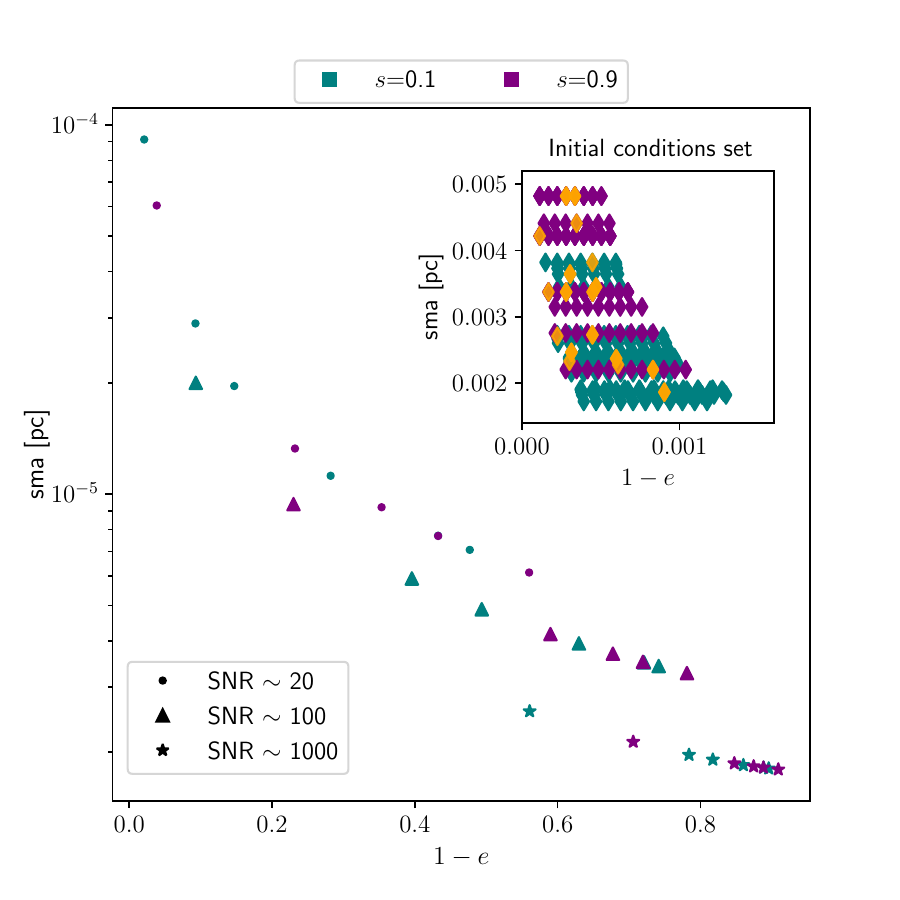}
    \caption{The orbital parameters of the XMRIs and their approximated SNR. The dot, triangle, and star markers on the main plot show XMRIs evolved from the initial conditions highlighted in orange in the upper right subplot, where we show a set of initial parameters obtained for both spin magnitudes s=0.1 (teal color) and s=0.9 (purple color). We use the systems with $\mathrm{SNR}\approx$20 and $\mathrm{SNR}\approx 100$ for the accuracy analysis.}
    \label{fig:smaecc}
\end{figure}

In figure~\ref{fig:smaecc}, we show the orbital parameters of the sources used for the accuracy analysis. The subplot in the upper right corner shows the set of initial orbital parameters with which the XMRIs are formed, considering the two different spin magnitudes, where the subset of 20 sources that are evolved up to the predefined SNR values is highlighted in orange. We also plot the loud sources with $\mathrm{SNR}\approx 1000$ in Figure~\ref{fig:smaecc}; nevertheless, we exclude these loud sources from the accuracy analysis as detecting even one of these will provide an accurate value when performing a parameter extraction, and it will not significantly benefit from detecting several sources.

\begin{figure}
    \centering
    \includegraphics[width=\linewidth]{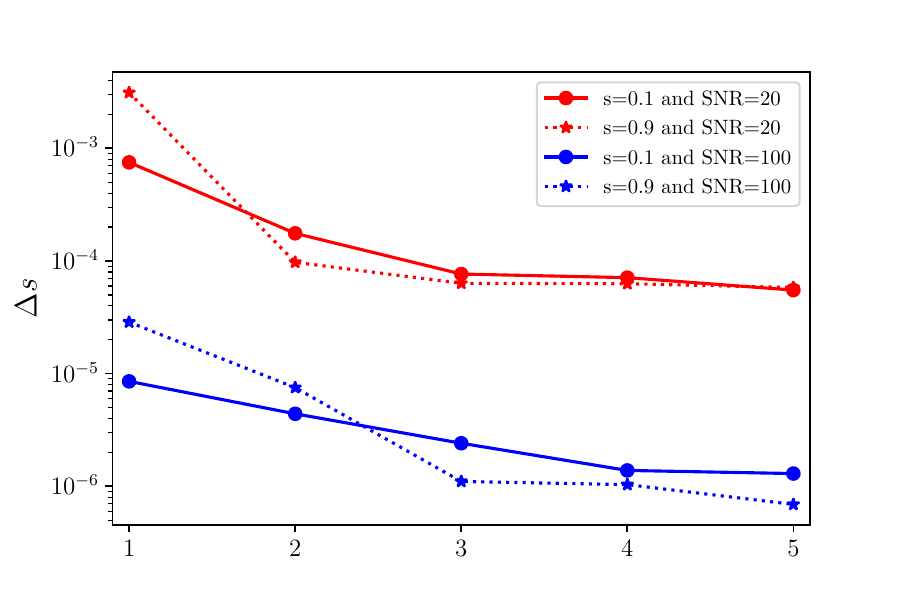}
    \includegraphics[width=\linewidth]{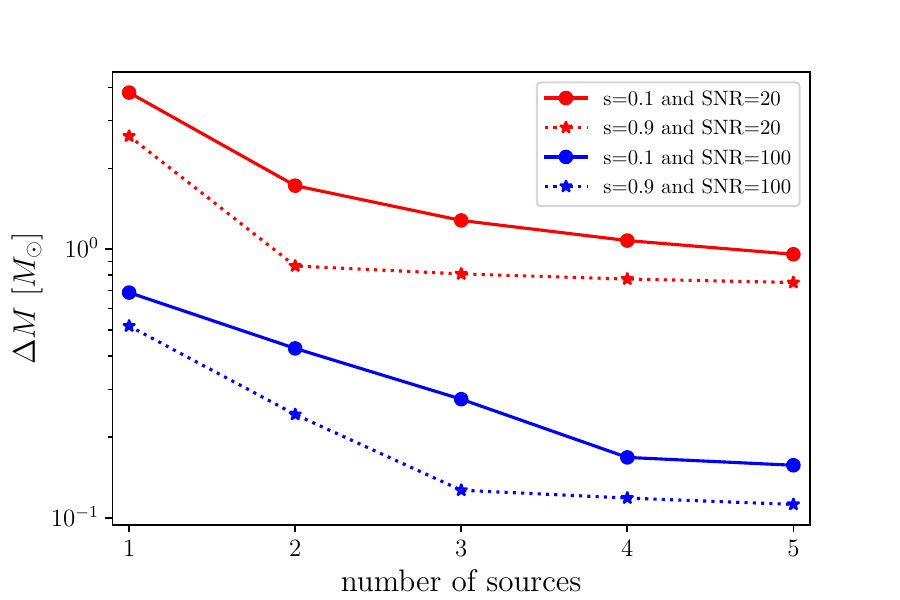}
    \caption{The accuracy of measuring the spin $s$ (upper panel) and the mass $M$ (lower panel) of SgrA$^\ast$ as a function of number of detected XMRIs.}
    \label{fig:para}
\end{figure}

Figure~\ref{fig:para} shows the detection error for the spin $\Delta s$ and the mass $\Delta M$ of SgrA$^\ast$ as the number of sources $N$ increases. As expected, sources with higher SNR provide higher accuracy than sources with lower SNR. We see that the accuracy is enhanced as $N$ grows in all cases; however, the spin accuracy is the quantity that benefits the most from detecting several sources. For low SNR ($\approx 20$), the accuracy of the spin increases one order of magnitude for $N=3$, and there is already an important improvement with just two sources, especially for the highly spinning case. After that, increasing the number of sources does not significantly improve the accuracy; it remains nearly constant and takes roughly the same value, $\Delta s \approx 10^{-4}$, independently of the spin magnitude. In the case of sources with SNR $\approx 100$, the highly spinning case, $s=0.9$, benefits the most from detecting several sources; when $N=5$, the accuracy $\Delta s$ increases two orders of magnitude, reaching $\Delta s \approx 10^{-7}$. Meanwhile, the accuracy for the slow spinning case, $s=0.1$, gets better by almost an order of magnitude for $N=5$.

The increasing number of sources also enhances the detection accuracy for the mass $\Delta M$. Systems with $s=0.9$ achieve a better mass estimation precision than the slow-spinning case, and for $N=5$, the accuracy is improved by roughly one order of magnitude in all cases. To compare the contribution of each XMRI to the total accuracy, we show in Table~\ref{tab:delta_div}, the ratio between the accuracy obtained with a single XMRI to the total accuracy $(\Delta s)_{\rm tot}$ and $(\Delta M)_{\rm tot}$ obtained with the set of $N=5$ XMRIs. The XMRIs that improve the accuracy the most are not always the ones with the highest SNR; although a combination of $e$ and $r_{\rm p}$ is important, the position of the pericenter distance is the parameter that influences $\Delta s$ the most. On the other hand, the contribution of a single XMRI to the total accuracy $(\Delta M)_{\rm tot}$ does not vary too much from source to source, the ratio $\Delta M/(\Delta M)_{\rm tot}$ remains between 1.79 and 6.50, while in the case of the spin, this ratio can reach up to 53.7.

The specific results are tightly related to the specific orbital parameters of a source. Nevertheless, the individual contribution to the mass accuracy remains similar in all cases; therefore, a different population of XMRIs will likely produce similar results for the mass measurements. For the spin, the orbital inclination plays a crucial role, as the effects of spin are not symmetric. However, while different orbital parameters can alter the individual estimates, the order of magnitude and the information encoded in the measurements will generally remain consistent.

\begin{table}
    \centering
    \begin{tabular}{c|c|c|c|c}
        \hline
        $s$ & $\iota$ & SNR & $\Delta s/(\Delta s)_\mathrm{tot}$ & $\Delta M/(\Delta M)_\mathrm{tot}$ \\
        \hline
        \multirow{5}{*}{0.1}
        & 0.1 & 29.7 & 13.6 & 4.00 \\
        & 0.4 & 28.0 & 5.59 & 4.88 \\
        & 0.7 & 25.0 & 2.54 & 4.84 \\
        & 1.0 & 17.5 & 45.1 & 3.51 \\
        & 1.57& 11.0 & 1.62 & 4.67 \\
        \hline
        \multirow{5}{*}{0.9}
        & 0.1 & 30.0 & 53.7 & 3.50 \\
        & 0.4 & 28.2 & 4.08 & 6.22 \\
        & 0.7 & 24.3 & 1.77 & 3.75 \\
        & 1.0 & 17.6 & 9.87 & 4.06 \\
        & 1.57& 11.6 & 2.72 & 6.50 \\
        \hline
        \multirow{5}{*}{0.1}
        & 0.1 & 169  & 6.58 & 4.39 \\
        & 0.4 & 172  & 4.40 & 4.00 \\
        & 0.7 & 163  & 2.37 & 3.06 \\
        & 1.0 & 96   & 1.48 & 1.79 \\
        & 1.57& 73   & 3.58 & 3.34 \\
        \hline
        \multirow{5}{*}{0.9}
        & 0.1 & 152  & 41.5 & 4.61 \\
        & 0.4 & 148  & 17.5 & 5.01 \\
        & 0.7 & 149  & 1.66 & 2.98 \\
        & 1.0 & 105  & 4.29 & 2.98 \\
        & 1.57& 70   & 1.38 & 5.05 \\
        \hline
    \end{tabular}
    \caption{The contribution of each source in a system of five sources. The contribution is shown in the last two columns as the ratio of the accuracy of a single XMRI $\Delta\theta(\theta=s,M)$ to the accuracy of a system of five XMRIs with similar SNRs $(\Delta\theta)_{\mathrm{tot}}$.}
    \label{tab:delta_div}
\end{table}

In the low-spin regime ($s = 0.1$), a single source with $\mathrm{SNR}\simeq 100$ achieves a precision in both spin $\Delta s$ and mass $\Delta M$ that is comparable, but still better, to that from five sources with $\mathrm{SNR}\simeq 20$. In this case, a single high-SNR detection would provide more accurate information than a small population of low-SNR sources. However, in the high-spin regime ($s = 0.9$), the performance gap narrows. Just two sources with $\mathrm{SNR} \simeq 20$ can achieve a spin and mass estimation accuracy of the same order of magnitude as a single $\mathrm{SNR} \simeq 100$ source. Our results indicate that the most probable scenario for XMRIs around SgrA$^\ast$ includes a population of XMRIs with $\mathrm{SNR}\in (10, 100)$, as roughly 80$\%$ of detectable sources fall in this interval. Multiple moderate-SNR events are statistically more likely than a single very high-SNR detection, especially considering the long inspiral timescales of XMRIs and the wide range of initial orbital parameters.

The red and blue lines in Figure~\ref{fig:para} provide two limits for parameter estimation accuracy, one for scenarios where all sources have relatively low $\mathrm{SNR}\simeq 20$, represented by the red lines, and the optimal case where all sources have high $\mathrm{SNR}\simeq 100$, represented by the blue lines. Therefore, in the conservative case where there are only five signals clearly detected, the accuracy of the resulting parameter estimations will be between these two extremes; the specific outcome depends on the actual SNR distribution of the detected sample, and even if these sources have a relatively low SNR, $\mathrm{SNR}\sim 20$, the accuracy achieved can be similar to that of a single sources with $\mathrm{SNR}\sim 100$.

\section{Conclusions}
\label{sec:Conclusions}

By performing a Fisher matrix analysis, we predict the accuracy with which the mass and spin of SgrA$^\ast$, can be determined. We have shown that even when the sources' SNR is relatively low ($\approx$ 20), the accuracy of the parameter extraction can be significantly enhanced by detecting more than one source. For the slowly spinning case $s=0.1$, we get that the spin accuracy $\Delta s$ is around one order of magnitude better when using five sources for the analysis, while for the highly spinning case $s=0.9$, the accuracy improves by even two orders of magnitude. The improvement of using multiple sources for the detection accuracy of the mass $\Delta M$ is around one order of magnitude, independent of the spin. How much adding one source to the analysis contributes to improving the accuracy depends on the properties of the inspiraling systems. For the spin, the improvement of its accuracy $\Delta s$ varies from 1.62 to 45.1 when adding one source to the analysis in the case of $s=0.1$, and from 1.77 to up to 53.7 for $s=0.9$. The accuracy of the mass $\Delta M$ improves, on average, by around 4 regardless of the value of the spin and the orbital parameters of the XMRI. 

We estimate the total number of sources inspiraling towards SgrA$^\ast$ for two different spin values. For a spin of $s=0.9$, we obtain $N_{\rm tot} = 16^{+3}_{-5}$ and for a spin of $s=0.1$, we obtain $N_{\rm tot} = 9^{+6}_{-4}$. From these sources, nearly 80\,\% have an SNR of $\approx10$-$100$ and around 20\,\% an $\mathrm{SNR}\approx100$-$1000$. These results are not significantly different from previous findings~\citep{Pau_2019, VVA_2023}. However, while previous estimates suggest that XMRIs orbiting SgrA$^\ast$ are highly eccentric, we find that the eccentricity of detectable systems ranges from 0.5 to 0.97 for XMRIs with $\mathrm{SNR}\approx10$-$100$ and from 0.2 to 0.92 for systems with $\mathrm{SNR}\approx100$-$1000$; indicating that some of the detectable XMRIs that might be orbiting SgrA$^\ast$ at the moment can have moderate eccentricity values. Additionally, we identify loud sources with $\mathrm{SNR}\gtrsim1000$ and eccentricities of around $e\approx 0.7$. However, the expected number of these sources remains low, representing only $\lesssim2\,\%$ of the total number of detectable sources orbiting SgrA$^\ast$.

Our sample represents a randomly drawn set of sources, and we expect these results to hold under different orbital parameter distributions. As discussed in Section~\ref{sec:Accuracy} , the contribution to mass estimation remains roughly the same across the sample, while the spin estimation is more sensitive to orbital configuration and source orientation. We find that for the low spinning scenario ($s = 0.1$), a single loud source with a $\mathrm{SNR} \simeq 100$ would provide measurements of the mass and spin of SgrA$^\ast$ that are approximately an order of magnitude better than those obtained from five sources with $\mathrm{SNR}\simeq 20$. In contrast, when we consider a high spin, $s=0.9$, detecting just two XMRIs with a $\mathrm{SNR} \simeq 20$ provides spin and mass estimates with an accuracy comparable to the one achieved with a single $\mathrm{SNR} \simeq 100$ source. In reality, the detectable population of XMRIs will likely consist of a mixture of sources with different SNR, with high $\mathrm{SNR} \gtrsim 100$ sources less common than moderate $\mathrm{SNR} \in (10,100)$ sources; nevertheless, the signal from a few low SNR sources can still provide information about the systems with an accuracy comparable to that from a single high SNR source.

An increased number of detected XMRIs improves the spin and mass estimates of SgrA$^\ast$, demonstrating the value of multi-source analysis. While the characteristics of gravitational wave sources in nature cannot be controlled, having prior knowledge of likely orbital parameters and SNRs helps refine search techniques and develop more accurate waveform models. These models are essential for detecting signals in the data expected from future space-based gravitational wave detectors. Thus, these eccentricity ranges should be considered when searching for XMRIs signals.

\section*{Acknowledgements}

We thank Pau Amaro-Seoane and Xian Chen for their helpful comments and interesting discussions. VVA acknowledges support from the Boya postdoctoral fellowship program of Peking University. YL is supported by the National Science Foundation of China (Grant No. 11873022) and the National Key Research and Development Program of China (Grant No. 2021YFC2203002). ATO acknowledges support from the Key Laboratory of TianQin Project (Sun Yat-sen University), Ministry of Education (China).

\section*{Data Availability}
The data underlying this article will be shared on reasonable request to the corresponding author.


\small
\bibliographystyle{mnras}
\bibliography{xmrisbib}

\begin{thebibliography}{}
\makeatletter
\relax
\def\mn@urlcharsother{\let\do\@makeother \do\$\do\&\do\#\do\^\do\_\do\%\do\~}
\def\mn@doi{\begingroup\mn@urlcharsother \@ifnextchar [ {\mn@doi@}
  {\mn@doi@[]}}
\def\mn@doi@[#1]#2{\def\@tempa{#1}\ifx\@tempa\@empty \href
  {http://dx.doi.org/#2} {doi:#2}\else \href {http://dx.doi.org/#2} {#1}\fi
  \endgroup}
\def\mn@eprint#1#2{\mn@eprint@#1:#2::\@nil}
\def\mn@eprint@arXiv#1{\href {http://arxiv.org/abs/#1} {{\tt arXiv:#1}}}
\def\mn@eprint@dblp#1{\href {http://dblp.uni-trier.de/rec/bibtex/#1.xml}
  {dblp:#1}}
\def\mn@eprint@#1:#2:#3:#4\@nil{\def\@tempa {#1}\def\@tempb {#2}\def\@tempc
  {#3}\ifx \@tempc \@empty \let \@tempc \@tempb \let \@tempb \@tempa \fi \ifx
  \@tempb \@empty \def\@tempb {arXiv}\fi \@ifundefined
  {mn@eprint@\@tempb}{\@tempb:\@tempc}{\expandafter \expandafter \csname
  mn@eprint@\@tempb\endcsname \expandafter{\@tempc}}}

\bibitem[\protect\citeauthoryear{{Akiyama} \& {et al.}}{{Akiyama} \& {et
  al.}}{2022a}]{SgrA1_2022}
{Akiyama} K.,  {et al.} 2022a, \mn@doi [\apjl] {10.3847/2041-8213/ac6674},
  \href {https://ui.adsabs.harvard.edu/abs/2022ApJ...930L..12A} {930, L12}

\bibitem[\protect\citeauthoryear{{Akiyama} \& {et al.}}{{Akiyama} \& {et
  al.}}{2022b}]{SgrA2_2022}
{Akiyama} K.,  {et al.} 2022b, \mn@doi [\apjl] {10.3847/2041-8213/ac6675},
  \href {https://ui.adsabs.harvard.edu/abs/2022ApJ...930L..13A} {930, L13}

\bibitem[\protect\citeauthoryear{{Akiyama} \& {et al.}}{{Akiyama} \& {et
  al.}}{2022c}]{SgrA3_2022}
{Akiyama} K.,  {et al.} 2022c, \mn@doi [\apjl] {10.3847/2041-8213/ac6429},
  \href {https://ui.adsabs.harvard.edu/abs/2022ApJ...930L..14A} {930, L14}

\bibitem[\protect\citeauthoryear{{Akiyama} \& {et al.}}{{Akiyama} \& {et
  al.}}{2022d}]{SgrA4_2022}
{Akiyama} K.,  {et al.} 2022d, \mn@doi [\apjl] {10.3847/2041-8213/ac6736},
  \href {https://ui.adsabs.harvard.edu/abs/2022ApJ...930L..15A} {930, L15}

\bibitem[\protect\citeauthoryear{{Akiyama} \& {et al.}}{{Akiyama} \& {et
  al.}}{2022e}]{SgrA5_2022}
{Akiyama} K.,  {et al.} 2022e, \mn@doi [\apjl] {10.3847/2041-8213/ac6672},
  \href {https://ui.adsabs.harvard.edu/abs/2022ApJ...930L..16A} {930, L16}

\bibitem[\protect\citeauthoryear{{Akiyama} \& {et al.}}{{Akiyama} \& {et
  al.}}{2022f}]{SgrA6_2022}
{Akiyama} K.,  {et al.} 2022f, \mn@doi [\apjl] {10.3847/2041-8213/ac6756},
  \href {https://ui.adsabs.harvard.edu/abs/2022ApJ...930L..17A} {930, L17}

\bibitem[\protect\citeauthoryear{{Alexander} \& {Hopman}}{{Alexander} \&
  {Hopman}}{2009}]{AlexanderTal_2009}
{Alexander} T.,  {Hopman} C.,  2009, \mn@doi [\apj]
  {10.1088/0004-637X/697/2/1861}, \href
  {https://ui.adsabs.harvard.edu/abs/2009ApJ...697.1861A} {697, 1861}

\bibitem[\protect\citeauthoryear{{Amaro-Seoane}}{{Amaro-Seoane}}{2018}]{Pau_2018}
{Amaro-Seoane} P.,  2018, \mn@doi [Living Reviews in Relativity]
  {10.1007/s41114-018-0013-8}, \href
  {https://ui.adsabs.harvard.edu/abs/2018LRR....21....4A} {21, 4}

\bibitem[\protect\citeauthoryear{Amaro-Seoane}{Amaro-Seoane}{2019}]{Pau_2019}
Amaro-Seoane P.,  2019, \mn@doi [Phys. Rev. D] {10.1103/PhysRevD.99.123025},
  99, 123025

\bibitem[\protect\citeauthoryear{Amaro-Seoane}{Amaro-Seoane}{2020}]{Pau_2020}
Amaro-Seoane P.,  2020, The gravitational capture of compact objects by massive
  black holes (\mn@eprint {arXiv} {2011.03059})

\bibitem[\protect\citeauthoryear{{Amaro-Seoane} \& {Preto}}{{Amaro-Seoane} \&
  {Preto}}{2011}]{Amaro-Seoane_2011}
{Amaro-Seoane} P.,  {Preto} M.,  2011, \mn@doi [Classical and Quantum Gravity]
  {10.1088/0264-9381/28/9/094017}, \href
  {https://ui.adsabs.harvard.edu/abs/2011CQGra..28i4017A} {28, 094017}

\bibitem[\protect\citeauthoryear{{Amaro-Seoane} \& {et al.}}{{Amaro-Seoane} \&
  {et al.}}{2017}]{LISA_2017}
{Amaro-Seoane} P.,  {et al.} 2017, preprint, \href
  {http://adsabs.harvard.edu/abs/2017arXiv170200786A} {} (\mn@eprint {arXiv}
  {1702.00786})

\bibitem[\protect\citeauthoryear{Amaro-Seoane, Gair, Freitag
  et~al.}{Amaro-Seoane et~al.}{2007}]{Amaro_2007}
Amaro-Seoane P.,  Gair J.~R.,  Freitag M.,   et~al., 2007, \mn@doi [Classical
  and Quantum Gravity] {10.1088/0264-9381/24/17/R01}, 24, R113

\bibitem[\protect\citeauthoryear{{Amaro-Seoane}, {Sopuerta}  \&
  {Freitag}}{{Amaro-Seoane} et~al.}{2013}]{Pau_2013}
{Amaro-Seoane} P.,  {Sopuerta} C.~F.,   {Freitag} M.~D.,  2013, \mn@doi
  [\mnras] {10.1093/mnras/sts572}, \href
  {https://ui.adsabs.harvard.edu/abs/2013MNRAS.429.3155A} {429, 3155}

\bibitem[\protect\citeauthoryear{{Babak} et~al.,}{{Babak}
  et~al.}{2017}]{babak_gair_2017}
{Babak} S.,  et~al., 2017, \mn@doi [\prd] {10.1103/PhysRevD.95.103012}, \href
  {https://ui.adsabs.harvard.edu/abs/2017PhRvD..95j3012B} {95, 103012}

\bibitem[\protect\citeauthoryear{{Bahcall} \& {Wolf}}{{Bahcall} \&
  {Wolf}}{1977}]{BW_1977}
{Bahcall} J.~N.,  {Wolf} R.~A.,  1977, \mn@doi [\apj] {10.1086/155534}, \href
  {https://ui.adsabs.harvard.edu/abs/1977ApJ...216..883B} {216, 883}

\bibitem[\protect\citeauthoryear{Barack \& Cutler}{Barack \&
  Cutler}{2004}]{Barack:2003fp}
Barack L.,  Cutler C.,  2004, \mn@doi [Phys. Rev. D]
  {10.1103/PhysRevD.69.082005}, 69, 082005

\bibitem[\protect\citeauthoryear{{Baumgardt}, {Amaro-Seoane}  \&
  {Sch{\"o}del}}{{Baumgardt} et~al.}{2018}]{Baum_2018}
{Baumgardt} H.,  {Amaro-Seoane} P.,   {Sch{\"o}del} R.,  2018, \mn@doi [\aap]
  {10.1051/0004-6361/201730462}, \href
  {https://ui.adsabs.harvard.edu/abs/2018A&A...609A..28B} {609, A28}

\bibitem[\protect\citeauthoryear{{Binney} \& {Tremaine}}{{Binney} \&
  {Tremaine}}{1987}]{Binney_Tremaine_1987}
{Binney} J.,  {Tremaine} S.,  1987, {Galactic dynamics}.
Princeton Univ. Press

\bibitem[\protect\citeauthoryear{{Coe}}{{Coe}}{2009}]{coe_2009}
{Coe} D.,  2009, arXiv e-prints, \href
  {https://ui.adsabs.harvard.edu/abs/2009arXiv0906.4123C} {p. arXiv:0906.4123}

\bibitem[\protect\citeauthoryear{{Colpi} et~al.,}{{Colpi}
  et~al.}{2024}]{lisa_2024}
{Colpi} M.,  et~al., 2024, \mn@doi [arXiv e-prints]
  {10.48550/arXiv.2402.07571}, \href
  {https://ui.adsabs.harvard.edu/abs/2024arXiv240207571C} {p. arXiv:2402.07571}

\bibitem[\protect\citeauthoryear{{Daly}, {Donahue}, {O'Dea}, {Sebastian},
  {Haggard}  \& {Lu}}{{Daly} et~al.}{2024}]{Daly_2024}
{Daly} R.~A.,  {Donahue} M.,  {O'Dea} C.~P.,  {Sebastian} B.,  {Haggard} D.,
  {Lu} A.,  2024, \mn@doi [\mnras] {10.1093/mnras/stad3228}, \href
  {https://ui.adsabs.harvard.edu/abs/2024MNRAS.527..428D} {527, 428}

\bibitem[\protect\citeauthoryear{{Dokuchaev}}{{Dokuchaev}}{2023}]{Dokuchaev_2023}
{Dokuchaev} V.~I.,  2023, \mn@doi [Astronomy] {10.3390/astronomy2030010}, \href
  {https://ui.adsabs.harvard.edu/abs/2023Astro...2..141D} {2, 141}

\bibitem[\protect\citeauthoryear{{Fang} \& {Huang}}{{Fang} \&
  {Huang}}{2020}]{Fang_Huang_2020}
{Fang} Y.,  {Huang} Q.-G.,  2020, \mn@doi [\prd] {10.1103/PhysRevD.102.104002},
  \href {https://ui.adsabs.harvard.edu/abs/2020PhRvD.102j4002F} {102, 104002}

\bibitem[\protect\citeauthoryear{{Finn}}{{Finn}}{1992}]{finn_1992}
{Finn} L.~S.,  1992, \mn@doi [\prd] {10.1103/PhysRevD.46.5236}, \href
  {https://ui.adsabs.harvard.edu/abs/1992PhRvD..46.5236F} {46, 5236}

\bibitem[\protect\citeauthoryear{{Finn} \& {Thorne}}{{Finn} \&
  {Thorne}}{2000}]{finn_thorne_2000}
{Finn} L.~S.,  {Thorne} K.~S.,  2000, \mn@doi [\prd]
  {10.1103/PhysRevD.62.124021}, \href
  {https://ui.adsabs.harvard.edu/abs/2000PhRvD..62l4021F} {62, 124021}

\bibitem[\protect\citeauthoryear{{Fragione} \& {Loeb}}{{Fragione} \&
  {Loeb}}{2020}]{Fragione_2020}
{Fragione} G.,  {Loeb} A.,  2020, \mn@doi [\apjl] {10.3847/2041-8213/abb9b4},
  \href {https://ui.adsabs.harvard.edu/abs/2020ApJ...901L..32F} {901, L32}

\bibitem[\protect\citeauthoryear{{Fragione} \& {Loeb}}{{Fragione} \&
  {Loeb}}{2022}]{Fragione_2022}
{Fragione} G.,  {Loeb} A.,  2022, \mn@doi [\apjl] {10.3847/2041-8213/ac76ca},
  \href {https://ui.adsabs.harvard.edu/abs/2022ApJ...932L..17F} {932, L17}

\bibitem[\protect\citeauthoryear{{Freitag}, {Amaro-Seoane}  \&
  {Kalogera}}{{Freitag} et~al.}{2006}]{Marc_2006}
{Freitag} M.,  {Amaro-Seoane} P.,   {Kalogera} V.,  2006, in Journal of Physics
  Conference Series. pp 252--258 (\mn@eprint {arXiv} {astro-ph/0607001}),
  \mn@doi{10.1088/1742-6596/54/1/040}

\bibitem[\protect\citeauthoryear{{Gallego-Cano}, {Sch{\"o}del}, {Dong},
  {Nogueras-Lara}, {Gallego-Calvente}, {Amaro-Seoane}  \&
  {Baumgardt}}{{Gallego-Cano} et~al.}{2018}]{Gallego-Cano_2018}
{Gallego-Cano} E.,  {Sch{\"o}del} R.,  {Dong} H.,  {Nogueras-Lara} F.,
  {Gallego-Calvente} A.~T.,  {Amaro-Seoane} P.,   {Baumgardt} H.,  2018,
  \mn@doi [\aap] {10.1051/0004-6361/201730451}, \href
  {https://ui.adsabs.harvard.edu/abs/2018A&A...609A..26G} {609, A26}

\bibitem[\protect\citeauthoryear{{Ghez} et~al.,}{{Ghez}
  et~al.}{2008}]{Ghez_2008}
{Ghez} A.~M.,  et~al., 2008, \mn@doi [\apj] {10.1086/592738}, \href
  {https://ui.adsabs.harvard.edu/abs/2008ApJ...689.1044G} {689, 1044}

\bibitem[\protect\citeauthoryear{{Gillessen}, {Eisenhauer}, {Trippe},
  {Alexander}, {Genzel}, {Martins}  \& {Ott}}{{Gillessen}
  et~al.}{2009}]{Gillessen_2009}
{Gillessen} S.,  {Eisenhauer} F.,  {Trippe} S.,  {Alexander} T.,  {Genzel} R.,
  {Martins} F.,   {Ott} T.,  2009, \mn@doi [\apj]
  {10.1088/0004-637X/692/2/1075}, \href
  {https://ui.adsabs.harvard.edu/abs/2009ApJ...692.1075G} {692, 1075}

\bibitem[\protect\citeauthoryear{{Gourgoulhon}, {Le Tiec}, {Vincent}  \&
  {Warburton}}{{Gourgoulhon} et~al.}{2019}]{gourgoulhon_le-tiec_2019}
{Gourgoulhon} E.,  {Le Tiec} A.,  {Vincent} F.~H.,   {Warburton} N.,  2019,
  \mn@doi [\aap] {10.1051/0004-6361/201935406}, \href
  {https://ui.adsabs.harvard.edu/abs/2019A&A...627A..92G} {627, A92}

\bibitem[\protect\citeauthoryear{{Gravity Collaboration} et~al.,}{{Gravity
  Collaboration} et~al.}{2024}]{GravityColl_2024}
{Gravity Collaboration} et~al., 2024, \mn@doi [\aap]
  {10.1051/0004-6361/202452274}, \href
  {https://ui.adsabs.harvard.edu/abs/2024A&A...692A.242G} {692, A242}

\bibitem[\protect\citeauthoryear{{Hopman} \& {Alexander}}{{Hopman} \&
  {Alexander}}{2005}]{Hopman_2005}
{Hopman} C.,  {Alexander} T.,  2005, \mn@doi [\apj] {10.1086/431475}, \href
  {https://ui.adsabs.harvard.edu/abs/2005ApJ...629..362H} {629, 362}

\bibitem[\protect\citeauthoryear{Hopman \& Alexander}{Hopman \&
  Alexander}{2006}]{Hopman_2006}
Hopman C.,  Alexander T.,  2006, \mn@doi [The Astrophysical Journal]
  {10.1086/506273}, 645, L133–L136

\bibitem[\protect\citeauthoryear{{Isoyama}, {Nakano}  \& {Nakamura}}{{Isoyama}
  et~al.}{2018}]{isoyama_nakano_2018}
{Isoyama} S.,  {Nakano} H.,   {Nakamura} T.,  2018, \mn@doi [Progress of
  Theoretical and Experimental Physics] {10.1093/ptep/pty078}, \href
  {https://ui.adsabs.harvard.edu/abs/2018PTEP.2018g3E01I} {2018, 073E01}

\bibitem[\protect\citeauthoryear{{Klein} et~al.,}{{Klein}
  et~al.}{2016}]{klein_barausse_2016}
{Klein} A.,  et~al., 2016, \mn@doi [\prd] {10.1103/PhysRevD.93.024003}, \href
  {https://ui.adsabs.harvard.edu/abs/2016PhRvD..93b4003K} {93, 024003}

\bibitem[\protect\citeauthoryear{{Kroupa}}{{Kroupa}}{2001}]{Kroupa_2001}
{Kroupa} P.,  2001, \mn@doi [\mnras] {10.1046/j.1365-8711.2001.04022.x}, \href
  {https://ui.adsabs.harvard.edu/abs/2001MNRAS.322..231K} {322, 231}

\bibitem[\protect\citeauthoryear{Li}{Li}{2013}]{li_2013}
Li T. G.~F.,  2013, PhD thesis, Vrije Universiteit Amsterdam

\bibitem[\protect\citeauthoryear{{Li} et~al.,}{{Li}
  et~al.}{2024}]{tianqin_2024}
{Li} E.-K.,  et~al., 2024, \mn@doi [arXiv e-prints]
  {10.48550/arXiv.2409.19665}, \href
  {https://ui.adsabs.harvard.edu/abs/2024arXiv240919665L} {p. arXiv:2409.19665}

\bibitem[\protect\citeauthoryear{{Luo} et~al.,}{{Luo}
  et~al.}{2016}]{TianQin_2016}
{Luo} J.,  et~al., 2016, \mn@doi [Classical and Quantum Gravity]
  {10.1088/0264-9381/33/3/035010}, \href
  {https://ui.adsabs.harvard.edu/abs/2016CQGra..33c5010L} {33, 035010}

\bibitem[\protect\citeauthoryear{{Mei} \& {et al.}}{{Mei} \& {et
  al.}}{2021}]{TianQin_2021}
{Mei} J.,  {et al.} 2021, \mn@doi [Progress of Theoretical and Experimental
  Physics] {10.1093/ptep/ptaa114}, \href
  {https://ui.adsabs.harvard.edu/abs/2021PTEP.2021eA107M} {2021, 05A107}

\bibitem[\protect\citeauthoryear{Peters}{Peters}{1964}]{Peters:1964}
Peters P.~C.,  1964, \mn@doi [Phys. Rev.] {10.1103/PhysRev.136.B1224}, 136,
  B1224

\bibitem[\protect\citeauthoryear{{Preto} \& {Amaro-Seoane}}{{Preto} \&
  {Amaro-Seoane}}{2010}]{Preto_2010}
{Preto} M.,  {Amaro-Seoane} P.,  2010, \mn@doi [\apjl]
  {10.1088/2041-8205/708/1/L42}, \href
  {https://ui.adsabs.harvard.edu/abs/2010ApJ...708L..42P} {708, L42}

\bibitem[\protect\citeauthoryear{{Sch{\"o}del}, {Gallego-Cano}, {Dong},
  {Nogueras-Lara}, {Gallego-Calvente}, {Amaro-Seoane}  \&
  {Baumgardt}}{{Sch{\"o}del} et~al.}{2018}]{Schodel_2018}
{Sch{\"o}del} R.,  {Gallego-Cano} E.,  {Dong} H.,  {Nogueras-Lara} F.,
  {Gallego-Calvente} A.~T.,  {Amaro-Seoane} P.,   {Baumgardt} H.,  2018,
  \mn@doi [\aap] {10.1051/0004-6361/201730452}, \href
  {https://ui.adsabs.harvard.edu/abs/2018A&A...609A..27S} {609, A27}

\bibitem[\protect\citeauthoryear{{Shapiro} \& {Teukolsky}}{{Shapiro} \&
  {Teukolsky}}{1983}]{shapiro_teukolsky_1983}
{Shapiro} S.~L.,  {Teukolsky} S.~A.,  1983, {Black holes, white dwarfs, and
  neutron stars : the physics of compact objects}.
Wiley-VCH

\bibitem[\protect\citeauthoryear{Sigurdsson \& Rees}{Sigurdsson \&
  Rees}{1997}]{Sigurdsson_1997}
Sigurdsson S.,  Rees M.~J.,  1997, \mn@doi [Monthly Notices of the Royal
  Astronomical Society] {10.1093/mnras/284.2.318}, 284, 318

\bibitem[\protect\citeauthoryear{{Sorahana}, {Yamamura}  \&
  {Murakami}}{{Sorahana} et~al.}{2013}]{Sorahana_2013}
{Sorahana} S.,  {Yamamura} I.,   {Murakami} H.,  2013, \mn@doi [\apj]
  {10.1088/0004-637X/767/1/77}, \href
  {https://ui.adsabs.harvard.edu/abs/2013ApJ...767...77S} {767, 77}

\bibitem[\protect\citeauthoryear{{Torres-Orjuela}, {Huang}, {Liang}, {Liu},
  {Wang}, {Ye}, {Hu}  \& {Mei}}{{Torres-Orjuela}
  et~al.}{2024}]{torres-orjuela_huang_2023}
{Torres-Orjuela} A.,  {Huang} S.-J.,  {Liang} Z.-C.,  {Liu} S.,  {Wang} H.-T.,
  {Ye} C.-Q.,  {Hu} Y.-M.,   {Mei} J.,  2024, \mn@doi [Science China Physics,
  Mechanics, and Astronomy] {10.1007/s11433-023-2308-x}, \href
  {https://ui.adsabs.harvard.edu/abs/2024SCPMA..6759511T} {67, 259511}

\bibitem[\protect\citeauthoryear{{V{\'a}zquez-Aceves}, {Lin}  \&
  {Torres-Orjuela}}{{V{\'a}zquez-Aceves} et~al.}{2023}]{VVA_2023}
{V{\'a}zquez-Aceves} V.,  {Lin} Y.,   {Torres-Orjuela} A.,  2023, \mn@doi
  [\apj] {10.3847/1538-4357/acde51}, \href
  {https://ui.adsabs.harvard.edu/abs/2023ApJ...952..139V} {952, 139}

\bibitem[\protect\citeauthoryear{{Ye}, {Fan}, {Torres-Orjuela}, {Zhang}  \&
  {Hu}}{{Ye} et~al.}{2024}]{ye_fan_2023}
{Ye} C.-Q.,  {Fan} H.-M.,  {Torres-Orjuela} A.,  {Zhang} J.-d.,   {Hu} Y.-M.,
  2024, \mn@doi [\prd] {10.1103/PhysRevD.109.124034}, \href
  {https://ui.adsabs.harvard.edu/abs/2024PhRvD.109l4034Y} {109, 124034}

\makeatother
\end{thebibliography}
\normalsize

\end{document}